\documentstyle[aps,amssymb,preprint,tighten,pra]{revtex}

\begin{document}
\draft
\title{\LARGE {\bf Generating Fock states and two-Fock states superposition from
circular states, in a trapped ion} \thanks{Proceedings of the VII-th
      International Conference on Squeezed States and Uncertainty Relations,
      Boston, MA, USA, 04-08 June, 2001,
      Eds. D. Han, Y.S. Kim and A. Sergienko,
      NASA Conference Publication.}}
\author{Salomon S. Mizrahi\thanks{E-mail: salomon@df.ufscar.br}}
\address{Departamento de F\'{\i}sica, CCET, Universidade Federal de S\~ao Carlos \\
Via Washington Luiz, km 235, S\~ao Carlos, 13565-905, SP, Brazil}
\author{Wagner D. Jos\'e}
\address{Departamento de Ci\^encias Exatas e Tecnol\'ogicas, Universidade Estadual de\\
Santa Cruz \\ Rod. Ilh\'eus-Itabuna, km 16, Ilh\'eus, BA, Brazil}
\maketitle

\begin{abstract}
We propose three schemes to engineer $2^M$ and $M+1$ circular states for the
motion of the center of mass of a trapped ion, $M$ being the number of laser
pulses. Since the ion is subjected to several laser pulses, we analyze the
necessary duration of each one for generating the circular states, and from
these, the Fock states and superposition of two-Fock states. We also
calculate the probability for obtaining the required states.
\end{abstract}


\vspace{8mm} %

\section{Introduction}

%
An exciting problem in Quantum Mechanics consists in proposing and then
generating experimentally states that do not exist in the natural world.
Recent advances in quantum optics and atoms manipulation in traps have
allowed to realize this challenge with much success. Many nonclassical
states have been proposed \cite
{yur86,dav92,vas93,jan95,jan96,zg97,gerry97,laweber96,ragi00} and some have
already been produced in QED superconducting cavities and in trapped atoms
\cite{meek96} and ions \cite{resf}.

Differently from experiments involving atoms interacting with
superconducting cavities, trapped ions are loosely coupled to the
environment, so decoherence is less manifest, being therefore the best
candidates to engineer states or to construct devices as logic gates for
implementing quantum computation. However, {\em decoherence times must be of
the order of 1ms} \cite{roos99}.

Along the line of quantum states engineering Law and Eberly \cite{laweber96}
proposed a model where an arbitrary field state in a cavity may be generated
by manipulating an atom (source) inside the cavity during the atom-field
interaction process, thus all target states (superpositions of Fock number
state) can be created from the same initial state. Matos Filho and Vogel
\cite{mv196} proposed a quantum nondemolition measurement of the motional
energy of a trapped atom confined in a harmonic potential; by monitoring the
interaction times of the lasers interacting with the two-level electronic
state of the atom permit to collapse any initial vibrational state into a
Fock state, however the Fock state in which the atom collapses after a
complete measurement sequence is not predictable. In Refs. \cite
{dav96,kneerlaw98} the authors proposed a general method to entangle quantum
states between electronic and vibrational degrees of freedom in order to
engineer a vibrational (trapped ion) target state out from an initial vacuum
one; they also analyzed the sensitivity of the method due to errors in the
amplitude and phase of the lasers. Simulation for preparation of Fock states
by observation of quantum jumps were reported in Ref. \cite{cir93}. See also
\cite{drobny98} for the synthesis of an arbitrary two-mode bosonic state.

More subtly, Fock states can also be obtained by a superposition of coherent
states evenly distributed on a circle in the phase space called {\em %
circular states}. Experimental schemes were proposed to generate such
superpositions as states of the electromagnetic (EM) field in
superconducting cavities \cite{vas93,jan96,zg97,gerry97,bas99} and also as
states of the harmonic motion of the center of mass (COM) of a trapped ion
\cite{zg98,wagner00}. Recently, a procedure to create an arbitrary coherent
states superposition of the vibrational motion of a trapped ion, localized
along a line in phase space representation (target state) was proposed \cite
{moya99} .

Here we analyze three schemes to engineer circular states (CS) as target
states for the COM of a trapped ion, estimate the total time of laser pulses
and the probability to produce such states. This contribution is organized
as follows: In section II we present a brief revision of the circular states
and establish the conditions to obtain as a result of the $N$ coherent
states interference in phase space: the vacuum state, Fock states, special
superpositions of two-Fock states. In section III we review the ion-laser
interaction process. In section IV we estimate the necessary duration of the
laser pulses to generate the wanted states and the probability for producing
them. Finally in section V we present a summary and conclusions. %

\section{Circular states, Fock states and two-Fock states}

%
The circular state, proposed and studied by Janszky, coworkers \cite
{jan95,jan96,jan93} and Gagen \cite{gag95}, generalizes the Schr\"{o}dinger
cat state. It consists of a superposition of $N$ coherent states $|\alpha_k
\rangle$ equally likely distributed on a circle of radius $r=|\alpha_k|$ in
the phase space defined by the c-number $\alpha_k$,
\begin{equation}
\left| \Psi _{N}(\alpha _{0})\right\rangle =\lambda
_{N}^{-1/2}\sum_{k=1}^{N}C_{k}\left| \alpha _{k}\right\rangle \qquad {\rm %
with}\qquad \alpha _{k}=\alpha _{0}e^{i\frac{2\pi k}{N}},
\label{circularstate}
\end{equation}
where $\lambda _{N}^{-1/2}$ is the normalization constant. For N even and $%
C_{k+N/2} =\pm C_{k}$, the circular state is even or odd:
\mbox{$ \left| \Psi_{N}(-\alpha _0) \right\rangle =\pm \left|
\Psi _N (\alpha _0)\right\rangle$} and $\hat{a}^{N}\left| \Psi _N (\alpha _0
) \right\rangle =\alpha _0^{N} \left| \Psi _N (\alpha _0 )\right\rangle $ .
Hereon we assume $N$ even and consider two kinds of CS's, an even and an
odd: \newline
\vspace{1mm} \noindent (a) For $C_{k}=1$, the circular state is even
\begin{equation}
\left| \Psi _{N}\right\rangle =\lambda _{N}^{-1/2}\sum_{k=1}^{N}\left|
\alpha _{k}\right\rangle ,  \label{psin}
\end{equation}
with normalization factor
\[
\lambda _{N}=\left[ N+2\sum_{k=1}^{N-1}ke^{-2r^{2}\sin ^{2}\left( \frac{\pi
}{N}k\right) }\cos \left[ r^{2}\sin \left( \frac{2\pi }{N}k\right) \right]
\right] \, .
\]
In the Fock states basis state (\ref{psin}) is
\[
\left| \Psi _{N}\right\rangle =Z_{N}^{-1/2}\sum_{n=0}^{\infty }\frac{\alpha
_{0}^{Nn}}{\sqrt{\left( Nn\right) !}}\left| Nn\right\rangle =\frac{\left|
0\right\rangle +\frac{r^{N}}{\sqrt{N!}}\left| N\right\rangle +\frac{r^{2N}}{%
\sqrt{\left( 2N\right) !}}\left| 2N\right\rangle +\cdot \cdot \cdot }{\sqrt{%
1+\frac{r^{2N}}{N!}+\frac{r^{4N}}{\left( 2N\right) !}+\cdot \cdot \cdot }},
\]
with normalization constant
\[
Z_{N}\left( r^{2}\right) =\sum_{n=0}^{\infty }\frac{\left( r^{2}\right) ^{Nn}%
}{\left( Nn\right) !}\, .
\]
Due to interference between the coherent states the circular states may
emulate essentially one Fock state or a superposition of two-Fock states
under the following conditions:
\[
\left\{
\begin{array}{l}
1\ll \left( e{r}^{2}/N\right) ^{N}\ll 4^{N} \quad \text{gives closely a Fock
state} \left| N\right\rangle \\
\left( er^{2}/N\right) ^{N}\ll 1 \quad \text{gives closely the vacuum state}
\left| 0\right\rangle \\
er^{2}/N\simeq 1 \quad \text{gives closely the superposition} \left|
0\right\rangle +\left| N\right\rangle \\
er^{2}/\left( 4N\right) \simeq 1 \quad \text{gives closely the superposition}
\left| N\right\rangle +\left| 2N\right\rangle.
\end{array}
\right.
\]

The probability for a circular state having exactly $n$ photons is
\begin{equation}
P_{n}\equiv \left| \left\langle n|\Psi _{N}(\alpha _{0})\right\rangle
\right| ^{2}=Z_{N}^{-1}\left( r^{2}\right) \frac{r^{2n}}{n!}\delta
_{n,Nk}=P_{Nk}\left( r^{2}\right)\,, (k=0,1,2,...;N=2,3,4...).  \label{probn}
\end{equation}
($N=1$ is trivially the coherent state) meaning that for fixed value of $N$,
the circular state has a finite probability for $0, N, 2N, 3N,...$ quanta
and zero probability for any other number of quanta. The normalization
constant $Z_{N}\left( r^{2}\right) $ is also the partition function of the
statistical distribution of quanta, thus the mean value of any power of the
number of quanta is given by
\[
\left\langle \hat{n}^{k}\right\rangle =\frac{1}{Z_{N}\left( y\right) }\left(
y\frac{\partial }{\partial y}\right) ^{k}Z_{N}\left( y\right) ,\ y=r^{2} \,
.
\]

(b) For $C_{k}=e^{i2\pi k/N}$ the circular state is odd,

\begin{equation}
\left| \tilde{\Psi}_{N}\right\rangle =\lambda _{N}^{-1/2}\sum_{k=1}^{N}e^{%
\frac{2\pi ik}{N}}\left| \alpha _{k}\right\rangle ,  \label{psin2}
\end{equation}
with normalization factor
\begin{equation}
\lambda _{N}=\left[ N+2\sum_{k=1}^{N-1}ke^{-2r^{2}\sin ^{2}\left( \frac{\pi
}{N}k\right) }\cos \left[ \frac{2\pi k}{N}+r^{2}\sin \left( \frac{2\pi }{N}%
k\right) \right] \right] \, .  \label{norm2}
\end{equation}
In terms of Fock states it can be written as
\begin{equation}
\left| \tilde{\Psi}_{N}\right\rangle =Z_{N}^{-1/2}\sum_{k=1}^{\infty }\frac{%
r^{(kN-1)}}{\sqrt{(kN-1)!}}\left| kN-1\right\rangle =\frac{\left|
N-1\right\rangle +r^{N}\left( \frac{\left( N-1\right) !}{\left( 2N-1\right) !%
}\right) ^{1/2}\left| 2N-1\right\rangle +...}{\sqrt{1+r^{2N}\frac{(N-1)!}{%
(2N-1)!}+...}},  \label{psin3}
\end{equation}
where the only Fock states present in the superposition are $%
N-1,2N-1,3N-1,...,kN-1$ $\left( k=1,2,...;N=2,3,...\right)$ and when $\left(
r^{2} e/4N\right)^{N} \ll 1$ only the first Fock state $\left|
N-1\right\rangle$ is important.

The partition function is
\[
Z _{N}\left(r^2\right)=\sum_{k=1}^{\infty }\frac{r^{2(kN-1)}}{(kN-1)!}\, ,
\]
and the probability for each Fock state $\left| n\right\rangle $ to be
present in the superposition is
\begin{equation}
P_{n}\left( r^{2}\right) \equiv \left| \left\langle n|\tilde{\Psi}%
_{N}\right\rangle \right| ^{2}=\lambda _{N}^{2}\frac{r^{2(kN-1)}}{(kN-1)!}%
\delta _{n,kN-1}=P_{kN-1}\left(r^2\right)\, .  \label{prob2}
\end{equation}

In Table 1 we present a summary of our results, where $a=r^2$, and for each
pair $(N,a)$, $P _{N,j}$ corresponds to the overlap between the Fock state
(or two-Fock states superposition) and the CS. $P_{\uparrow }(a)$ is the
probability for the experimental production of a particular state for the
ion COM vibrational motion. $\bar{n}$ and {Var}$(\hat{n})$ stand for the
mean numbers and variances of quanta. \vspace{3mm}

\begin{center}
\begin{tabular}{|l|l|l|l|l|l|}
\hline
{Special Even Circular States} & $(N,a)$ & $P_{N,j}$ & $P_{\uparrow }(a)$ & $%
\bar{n}$ & {Var}$(\hat{n})$ \\ \hline
$|\tilde{\Psi}_{8,0}\rangle \approx \left( |0\rangle +|8\rangle \right) /%
\sqrt{2}$ & $(8,3.76)$ & $0.999960$ & $0.04635$ & $4.00322$ & $16.00497$ \\
\hline
$|\tilde{\Psi}_{8}^{(1)}\rangle \approx |8\rangle $ & $(8,6.80)$ & $0.982674$
& $0.12851$ & $7.99994$ & $1.10908$ \\ \hline
$|\tilde{\Psi}_{8,1}\rangle \approx \left( |8\rangle +|16\rangle \right) /%
\sqrt{2}$ & $\left( 8,12.17\right) $ & $0.991848$ & $0.12016$ & $11.94293$ &
$16.99720$ \\ \hline
$|\tilde{\Psi}_{8}^{(2)}\rangle \approx |16\rangle $ & $(8,15.80)$ & $%
0.79002 $ & $0.12542$ & $15.99847$ & $13.62115$ \\ \hline
$|\tilde{\Psi}_{16,0}\rangle \approx \left( |0\rangle +|16\rangle \right) /%
\sqrt{2}$ & $(16,6.80)$ & $0.999999$ & $0.00261$ & $7.99560$ & $63.99998$ \\
\hline
$|\tilde{\Psi}_{16}^{(1)}\rangle \approx |16\rangle $ & $(16,12.80)$ & $%
0.999918$ & $0.08564$ & $16.00001$ & $0.02088$ \\ \hline
$|\tilde{\Psi}_{16,1}\rangle \approx \left( |16\rangle +|32\rangle \right) /%
\sqrt{2}$ & $\left( 16,24.10\right) $ & $0.999807$ & $0.04295$ & $24.12047$
& $64.05677$ \\ \hline
$|\tilde{\Psi}_{16}^{(2)}\rangle \approx |32\rangle $ & $(16,31.20)$ & $%
0.96834$ & $0.07190$ & $32.02314$ & $8.10407$ \\ \hline
\end{tabular}
\end{center}

\vspace{1mm}

Table 1. {\small The first column stands for the particular states coming
out from a N-even CS, due to interference in phase space; the second column
gives the corresponding pairs of values ($N,a$) ($a\equiv r^2$) to be
attributed to the CS; the third column stands for the maximum probability of
a particular state in the CS superposition ; the fourth column shows the
probability for the experimental production of a particular state for the
ion COM vibrational motion; finally, fifth and sixth columns give the
corresponding mean numbers and variances.} %

\section{Ion-Laser Interaction}

%
A trapped ion is considered moving in a one-dimensional harmonic effective
pseudopotential, interacting with two laser (frequencies $\omega _{1}$ and $%
\omega _{2}$) in a Raman-type configuration, which is responsible for a
forbidden transition between two metastable internal electronic states, $%
|\uparrow \rangle $ and $|\downarrow \rangle $, separated by frequency $%
\omega _{0}$, and called excited and ground, respectively. In this
configuration, the two levels are indirectly coupled via a third one, $|r
\rangle $, which is adiabatically eliminated. A fourth level $|d\rangle $ is
used to cool the ion and to measure its internal electronic state state by
fluorescence emission, moreover it can still be used to generate the ion
motional states.

The hamiltonian describing the effective interaction between the two-level
electronic states and the quantized motion of the ion COM is written as (we
consider $\hbar =1$):
\begin{equation}
H=\omega a^{\dagger}a +\frac{\delta }{2} \sigma _{z} - \Omega \left( \sigma
_{-} e^{- i\eta (a+a^{\dagger })+ i\phi} +\sigma _{+} e^{ i\eta
(a+a^{\dagger })- i\phi }\right) ,  \label{ham1}
\end{equation}
where $\sigma _{+}$($\sigma _{-}$)=$|\uparrow \rangle \langle \downarrow |$($%
|\downarrow \rangle \langle \uparrow |$) and $\sigma _{z}$ are the usual
Pauli pseudospin operators, $a^{\dagger }$ ($a$) is the creation
(annihilation) operator of vibrational quanta, $\Omega $ is the effective
Rabi frequency of the transition $|\uparrow \rangle \leftrightarrow
|\downarrow \rangle $, $\omega$ is the ion vibrational frequency, $\delta
=\omega_1 - \omega_2 - \omega_0$ is the detuning frequency and
\begin{equation}
\eta = \Delta k/{\sqrt{ 2m\omega }},  \label{eta}
\end{equation}
is the Lamb-Dicke parameter, where $\Delta k=(\vec{k}_{1}-\vec{k}_{2})\cdot
\hat{x}$ and $|\vec{k}_{1(2)}|=\omega _{1(2)}/c$, being $\vec{k}_{1}\left(
\vec{k}_{2}\right) $ the wave vector of laser $1(2)$, $\hat{x}$ is the
operator referring to the COM position of the ion, $m$ is its mass and $%
\omega $ is the angular frequency of its (very closely) harmonic oscillation
in the trap. Thus, the strength of $\eta$ can be arbitrarily selected by
changing the relative direction of the laser beams.

Writing first $H$ in the interaction picture and then expanding the
resulting hamiltonian in terms of the Lamb-Dicke parameter one gets \cite
{mv98}
\begin{equation}
H_{I}=- \Omega e^{-\eta ^{2}/2}\left[ \sum_{m,l=0}^{\infty }\frac{(i\eta
)^{m+l}}{m!l!}a^{{\dagger }^{m}}a^{l}e^{i\left[ (m-l)\omega +\delta \right]
t+i\phi }\sigma _{-}+h.c.\right] .  \label{hexp}
\end{equation}
Since the frequency $\omega $ is large in comparison with $\Omega $,
resonance conditions result for $\delta =-k\omega $ $(k=m-l)$.

\noindent (a) The choice $k=0$ selects the carrier transition, with
Hamiltonian
\[
H_{I}^{(c)}=- \Omega f_{0}(\hat{n})\left[ \sigma _{+}e^{-i(\Delta t-\phi
)}+\sigma _{-}e^{i(\Delta t-\phi )}\right]
\]
(b) $k>0$ selects the $k$-th red frequency, whose Hamiltonian is
\[
H_{I}^{(r)}=- \Omega \left[ f_{k}(\hat{n})\sigma _{+}a^{k}e^{-i(\Delta
t-\phi )}+f_{k}^{\dagger }(\hat{n})\left( a^{\dagger }\right) ^{k}\sigma
_{-}e^{i(\Delta t-\phi )}\right]
\]
(c) $k<0$ ($k^{\prime}=-k$) selects the $k$-th blue frequency, with
Hamiltonian
\[
H_{I,k^{\prime }}^{(b)}=- \Omega \left[ \left( a^{\dagger }\right)
^{k^{\prime }}f_{k^{\prime }}(\hat{n})\sigma _{+}e^{-i(\Delta t-\phi
)}+f_{k^{\prime }}^{\dagger }(\hat{n})a^{k^{\prime }}\sigma _{-}e^{i(\Delta
t-\phi )}\right]
\]
where $f_{k}(\hat{n})=e^{-\eta ^{2}/2}\sum_{l=0}^{\hat{n}}\frac{\left( i\eta
\right) ^{2l+k}}{l!\left( l+k\right) !}\frac{\hat{n}!}{\left( \hat{n}
-l\right) !}.$ So, engineering a particular quantum state becomes possible
by choosing a particular hamiltonian from Eq. (\ref{hexp}). Considering the
Lamb-Dicke limit, $\eta ^{2}\ll 1$, some specific hamiltonians have been
proposed and investigated, namely: the Jaynes-Cummings hamiltonian ($k=1$),
the anti-Jaynes-Cummings hamiltonian ($k^{\prime }=1$) and the carrier
hamiltonian ($k=0$), which couple, respectively, the levels $\left|
m+1\downarrow \right\rangle \longleftrightarrow \left| m\uparrow
\right\rangle $, $\left| m\downarrow \right\rangle \longleftrightarrow
\left| m+1\uparrow \right\rangle $ and $\left| m\downarrow \right\rangle
\longleftrightarrow \left| m\uparrow \right\rangle $.

The measurement of the ion vibrational state is achieved indirectly through
the measurement of the ion electronic state which is realized by collecting
the resonance fluorescence signal from the transition $|d\rangle
\leftrightarrow |\downarrow \rangle $ through another laser strongly coupled
to the electronic ground state. The measured signal is the probability of
the ion to be found in the internal state $\left| \downarrow \right\rangle $%
. Since the fluorescence emission disturbs the COM motion of the ion, for
each new measurement the ion should be cooled back to its ground state. %

\section{Generation of circular states in a trapped ion}

%
We propose and analyze three different schemes to generate circular states
of the kind of Eqs. (\ref{psin}) and (\ref{psin2}). For this purpose we
consider a type-Kerr interaction between the ion and the effective laser
which is realized by tuning it resonantly $\left( \delta =0\right) $ with
the electronic transition frequency $\omega _{0}$. In the Lamb-Dicke limit,
the carrier hamiltonian becomes
\begin{equation}
H = \Omega \sigma _{x}-\eta ^{2} \Omega \left[ \left( 1+\frac{\eta ^{2}}{4}%
\right) a^{\dagger }a-\frac{\eta ^{2}}{4}\left( a^{\dagger }a\right)
^{2}+O\left( \eta ^{4}\right) \right] \sigma _{x} \approx \Omega \sigma
_{x}-\eta ^{2} \Omega a^{\dagger }a\sigma _{x}\,,  \label{happrox}
\end{equation}
(setting $\phi =\pi $ without lost of generality). The approximation is
valid under the condition
\begin{equation}
\eta ^{2}\left( Q+\bar{n}+1\right)/4 \ll 1\,,  \label{cond}
\end{equation}
where, $Q\equiv (\mbox{Var}(\hat{n})-\bar{n})/ \bar{n}$ is the Mandel
Q-parameter, $\bar{n}$ is the mean number of the ion motional quanta and $%
\mbox{Var}(\hat{n})$ is the variance. For instance, if the ion is prepared
initially in a coherent state $|\alpha _{0}\rangle $, then (\ref{cond})
becomes $(|\alpha _{0}|^{2}+1)\eta ^{2}/4\ll 1$, which must be satisfied in
order to consider the approximation in Eq. (\ref{happrox}). Thus, assuming
condition (\ref{cond}) that allow the use of (\ref{happrox}), the evolution
operator is
\begin{equation}
{U(t)\equiv e}^{{-iHt}}{=e}^{{-i \Omega t\sigma }_{x}}{e}^{{i\bar{ \Omega}%
t\sigma }_{x}{\ a}^{\dagger }{a}},  \label{ut}
\end{equation}
where $\bar{ \Omega}\equiv \eta ^{2} \Omega $. In all three schemes the ion
is initially prepared in the upper electronic state $\left| \uparrow
\right\rangle $ and its COM is in a coherent state,
\begin{equation}
\left| \Psi (0)\right\rangle \equiv \left| \alpha _{0}\right\rangle \otimes
\left| \uparrow \right\rangle =\left| \alpha _{0}\right\rangle \otimes
\left( \frac{\left| \uparrow _{x}\right\rangle +\left| \downarrow
_{x}\right\rangle }{\sqrt{2}}\right) ,
\end{equation}
the second equality follows because in terms of the eigenvectors $\left|
\uparrow \right\rangle $, $\left| \downarrow \right\rangle $ of $\sigma _{z}$%
, one has $\left| \uparrow _{x}\right\rangle =\left( \left| \uparrow
\right\rangle +\left| \downarrow \right\rangle \right)/\sqrt{2}$ and $\left|
\downarrow _{x}\right\rangle =\left( \left| \uparrow \right\rangle -\left|
\downarrow \right\rangle \right)/ \sqrt{2} $.

The procedure is the following: one applies $M$ laser pulses, each with
duration $t_{k}$, chosen according to the state one wants to generate. The $M
$ pulses are triggered in sequence as long as one observes $M$
no-fluorescence events when measuring the $\left| d\right\rangle
\leftrightarrow \left| \downarrow \right\rangle $ transition. If in one of
the measurement the ion is found in the ground electronic state $\left|
\downarrow \right\rangle $ (fluorescence is observed) the process is stopped
and one should repeat the sequence of pulses after preparing again the ion
in a coherent state. The no-fluorescence measurement is a necessary
condition because it assures the no-recoil of the ion vibrational COM
motion, thus, maximizing the probability of this particular sequence be
realized successfully. %

\subsection{First Scheme - $2^{M}$ even circular states, with $M$ laser
pulses}

%
Our aim is to generate a circular state (\ref{psin}) with $N=2^{M}$ coherent
states. We have to perform two operations repeatedly on the ion with the
lasers, (1) evolve unitarily the ion state with $U(t)$ (\ref{ut}) and (2) do
a projection of the ion statevector on $\left| \uparrow \right\rangle $ in
order to measure the probability to be found in the upper electronic state.
So, in this scheme each cycle involves the application of one pulse and one
measurement. The first pulse having duration $t_{1}$, the ion state becomes
\begin{eqnarray}
\left| \Psi (t_{1})\right\rangle &=&U(t_{1})\left| \Psi (0)\right\rangle =
\frac{1}{2}\left( e^{-i \Omega t_{1}}\left| \alpha (t_{1})\right\rangle
+e^{i \Omega t_{1}}\left| \alpha (-t_{1})\right\rangle \right) \otimes
\left| \uparrow \right\rangle  \nonumber \\
&+&\frac{1}{2}\left( e^{-i \Omega t_{1}}\left| \alpha (t_{1})\right\rangle
-e^{i \Omega t_{1}}\left| \alpha (-t_{1})\right\rangle \right) \otimes
\left| \downarrow \right\rangle ,  \label{ps1}
\end{eqnarray}
where $\alpha (t)\rangle \equiv |\alpha _{0}e^{i\bar{ \Omega}t}\rangle $.
The evolution is followed by a fluorescence measurement, if the ion is found
in state $\left| \uparrow \right\rangle $ (no fluorescence) we consider it a
{\em successful} measurement, then state (\ref{ps1}) reduces to
\begin{equation}
\left| \Psi ^{\prime }(t_{1})\right\rangle =\frac{{\cal N}_{1}}{2}\left(
e^{-i \Omega t_{1}}\left| \alpha (t_{1})\right\rangle +e^{i \Omega
t_{1}}\left| \alpha (-t_{1})\right\rangle \right) \otimes \left| \uparrow
\right\rangle ,
\end{equation}
where ${\cal N}_{1}$ is the normalization factor. The probability to find
the ion in state $\left| \uparrow \right\rangle $ is
\begin{equation}
P_{\uparrow }(t_{1})=\frac{1}{2}\left( 1+e^{-r^{2}\left( 1-\cos \left[ 2\bar{%
\Omega}t_{1}\right] \right) }\cos \left[ r^{2}\sin \left[ 2\bar{ \Omega}%
t_{1}\right] -2 \Omega t_{1}\right] \right) .  \label{pupt1}
\end{equation}
Assuming the ion was measured in this state, a second laser pulse drives the
ion into state $\left| \Psi (t_{1}+t_{2})\right\rangle =U(t_{2})\left| \Psi
^{\prime }(t_{1})\right\rangle $ and the second successful measurement
reduces this state to
\begin{eqnarray}
\left| \Psi ^{\prime }(t_{1}+t_{2})\right\rangle &=&\frac{{\cal N}_{2}}{4}%
\left( e^{-i \Omega (t_{1}+t_{2})}\left| \alpha (t_{1}+t_{2})\right\rangle
+e^{-i \Omega (t_{1}-t_{2})}\left| \alpha (t_{1}-t_{2})\right\rangle \right.
\nonumber \\
&&\left. +e^{i \Omega (t_{1}-t_{2})}\left| \alpha (t_{2}-t_{1})\right\rangle
+e^{-i \Omega (t_{1}+t_{2})}\left| \alpha (-t_{1}-t_{2})\right\rangle
\right) \otimes \left| \uparrow \right\rangle ,  \label{ps2}
\end{eqnarray}
with probability
\begin{eqnarray}
P_{\uparrow }(t_{1}+t_{2}) &=&\frac{1}{4}\left( 1+e^{-r^{2}\left( 1-\cos
\left[ 2\bar{ \Omega}t_{1}\right] \right) }\cos \left[ r^{2}\sin \left[ 2%
\bar{ \Omega}t_{1}\right] -2 \Omega t_{1}\right] \right.  \nonumber \\
&&\left. +e^{-r^{2}\left( 1-\cos \left[ 2\bar{ \Omega}t_{2}\right] \right)
}\cos \left[ r^{2}\sin \left[ 2\bar{ \Omega}t_{2}\right] -2 \Omega
t_{2}\right] \right.  \nonumber \\
&&\left. +e^{-r^{2}\left( 1-\cos \left[ 2\bar{ \Omega}\left(
t_{1}+t_{2}\right) \right] \right) }\cos \left[ r^{2}\sin \left[ 2\bar{
\Omega }\left( t_{1}+t_{2}\right) \right] -2 \Omega \left(
t_{1}+t_{2}\right) \right] \right.  \nonumber \\
&&\left. +e^{-r^{2}\left( 1-\cos \left[ 2\bar{ \Omega}\left(
t_{1}-t_{2}\right) \right] \right) }\cos \left[ r^{2}\sin \left[ 2\bar{
\Omega }\left( t_{1}-t_{2}\right) \right] -2 \Omega \left(
t_{1}-t_{2}\right) \right] \right) ,  \label{pupt2}
\end{eqnarray}
and so on.

In order to engineer the ion vibrational state as Eq. (\ref{psin}) we need
to adjust the phases of the coherent states such to be evenly distributed on
the circle, this is possible when the duration of the $k-$th pulse is
\begin{equation}
t_{k}=\frac{\pi }{2^{k}\bar{ \Omega}}.  \label{tp1}
\end{equation}
Besides, we remind that all the coefficients of superposition (\ref{ps2}),
for instance, should be equal to 1, so choosing the Lamb-Dicke parameter as $%
\eta ^{2}=2^{-\left( M+1\right) }$, it becomes possible to generate the
target superposition state.

The probability to get the ion successfully in the upper electronic state
after each of the $M$ sequential measurements is given by
\begin{equation}
P_{\uparrow }\left( \sum_{k=1}^{M}t_{k}\right) =\frac{1}{2^{M}}\left\{ 1+%
\frac{1}{2^{M-1}}\sum_{k=1}^{2^{M}-1}k\exp \left[ -2r^{2}\sin ^{2}\left(
\frac{\pi k}{2^{M}}\right) \cos \left[ r^{2}\sin \left( \frac{\pi k}{2^{M-1}}%
\right) \right] \right] \right\} .
\end{equation}
In the coherent states $\left| \alpha _{k}\right\rangle $ the c-numbers $%
\alpha _{k}$ have the following phases distributed on the circle of radius $%
r $
\begin{equation}
\theta _{k}^{(\pm)}=\theta _{0}\pm \frac{\pi }{2^{M}}\left( 2k-1\right)
\qquad k=1,...,2^{M-1}.  \label{thk}
\end{equation}
For example, if one wants to engineer a superposition with $16$ coherent
states ($M=4$ is the number of cycles) we need $\eta \approx 0.18$.
Proceeding along the lines drawn above we get the following states after
each successful measurement:
\[
t_{1}=\frac{\pi }{2\bar{ \Omega}}\Rightarrow \left| \Psi ^{\prime }\left(
t_{1}\right) \right\rangle ={\cal N}_{1}\left( \left| \alpha _{0}e^{i\pi
/2}\right\rangle +\left| \alpha _{0}e^{-i\pi /2}\right\rangle \right) /2
\]
\[
t_{2}=\frac{\pi }{4\bar{ \Omega}}\Rightarrow \left| \Psi ^{\prime }\left( t_{%
\bar{1}}+t_{2}\right) \right\rangle ={\cal N}_{2}\left( \left| \alpha
_{0}e^{3i\pi /4}\right\rangle +\left| \alpha _{0}e^{i\pi /4}\right\rangle
+\left| \alpha _{0}e^{-i\pi /4}\right\rangle \left| \alpha _{0}e^{-3i\pi
/4}\right\rangle \right) /4
\]
\begin{eqnarray*}
t_{3} &=&\frac{\pi }{8\bar{ \Omega}}\Rightarrow \left| \Psi ^{\prime }\left(
t_{\bar{1}}+t_{2}+t_{3}\right) \right\rangle ={\cal N}_{3}\left[ \left|
\alpha _{0}e^{7i\pi /8}\right\rangle +\left| \alpha _{0}e^{5i\pi
/8}\right\rangle +\left| \alpha _{0}e^{3i\pi /8}\right\rangle +\left| \alpha
_{0}e^{i\pi /8}\right\rangle +\right. \\
&&\left. \left| \alpha _{0}e^{-i\pi /8}\right\rangle +\left| \alpha
_{0}e^{-3i\pi /8}\right\rangle +\left| \alpha _{0}e^{-5i\pi /8}\right\rangle
+\left| \alpha _{0}e^{-7i\pi /8}\right\rangle \right] /8 ,
\end{eqnarray*}
etc. So, it suffices to control the on-off switchings of the laser beams to
get the right state. The total time necessary for the $M$ pulses is
\begin{equation}
T=\sum_{k=1}^{M}t_{k}=\frac{\pi }{\eta ^{2} \Omega }\left( 1-\frac{1}{2^{M}}
\right) =\frac{2^{M+1}\pi }{ \Omega }(1-\frac{1}{2^{M}})=\frac{2\pi }{
\Omega } (2^{M}-1)\approx (2^{M}-1)\mu s,
\end{equation}
since $2\pi /\Omega \simeq 1\mu s$, so about $15\mu s$ is the required
duration of pulses to get a superposition of $16$ coherent states, this time
is much less than that necessary for doing the experiment proposed in \cite
{dav96}.

The probability for the first successful cycle is $P_{\uparrow
}(t_{1})=\left[ 1+e^{-2r^{2}}\right] /2$ while for the second successful
cycle is $P_{\uparrow }(t_{1}+t_{2})=\frac{1}{4}\left[
1+e^{-2r^{2}}+2e^{-r^{2}}\cos r^{2}\right] $ and so forth.

Since the times $t_{k}$ are already fixed we only have the freedom to choose
the radius $r$ (the intensity of initial coherent state) and the number of
pulses to engineer a particular state. An example: For $2^{4}=16$ superposed
coherent states and $r=3.6$ the probability for producing approximately a
Fock state $\left| 16\right\rangle $ is $\approx 0.09$, which is not a bad
result since in the average one in eleven runs is successful, in this case
both conditions $1 \ll \left( er^{2}/2^{4}\right) ^{2^{4}} \ll 4^{16}$ and $%
\quad r^2 \ll 8(16)-1 $ are satisfied. See the fourth column of Table 1 for
further results. %

\subsection{Second scheme - $N=M+1$ even circular states}

%
Now, we will describe a scheme that generates a superposition of $M+1$
arbitrary coherent states on the circle as a result of selecting all pulses
having the same duration $\tau $.

If we consider equal times $\tau $ in Eqs. (\ref{ps1}) and (\ref{ps2}),
after two successful cycles we get
\begin{eqnarray}
\left| \Psi ^{\prime }(\tau )\right\rangle &=&\frac{{\cal N}_{1}}{2}\left(
\left| \alpha (\tau )\right\rangle +\left| \alpha (-\tau )\right\rangle
\right) \otimes \left| \uparrow \right\rangle ,  \nonumber \\
\left| \Psi ^{\prime }(2\tau )\right\rangle &=&\frac{{\cal N}_{2}}{4}\left(
\left| \alpha (2\tau )\right\rangle +2\left| \alpha (0)\right\rangle +\left|
\alpha (-2\tau )\right\rangle \right) \otimes \left| \uparrow \right\rangle ,
\label{ps12}
\end{eqnarray}
which is not what we need because the second term has a factor $2$ and we
want all the coefficients in (\ref{ps12}) be equal to $1$. Thus, we have to
use a second pair of lasers (also in the Lamb-Dicke limit) with Lamb-Dicke
parameter $\eta _{r}$ satisfying the condition $\eta _{r}^{2}\Lambda \ll
\eta ^{2} \Omega $ in order to rotate the electronic levels according to the
evolution ruled by
\begin{equation}
H_{r}=\Lambda \sigma _{x}\Rightarrow U_{r}(t)=e^{-i\Lambda t\sigma _{x}}.
\end{equation}
Now, each cycle consists of one rotating pulse, one evolution pulse and one
measurement. The action of the $l-$th rotating pulse on $\left| \uparrow
\right\rangle $ with duration $t_{l}^{\prime }$ is
\begin{equation}
U_{r}(t_{l}^{\prime })\left| \uparrow \right\rangle =\frac{e^{-i\Lambda
t_{l}^{\prime }}}{\sqrt{2}}\left| \uparrow _{x}\right\rangle +\frac{%
e^{i\Lambda t_{l}^{\prime }}}{\sqrt{2}}\left| \downarrow _{x}\right\rangle =%
\frac{a_{l}}{\sqrt{2}}\left| \uparrow _{x}\right\rangle +\frac{b_{l}}{\sqrt{2%
}}\left| \downarrow _{x}\right\rangle \, .
\end{equation}
Assuming each cycle is successful (no-fluorescence measurement), the
evolution of the ion vibrational state after, for instance, $M=3$ cycles
goes as follow ($\Omega \tau = 2 \pi$):
\begin{eqnarray}
&&\left| \alpha \right\rangle \rightarrow a_{1}\left| \alpha (\tau
)\right\rangle +b_{1}\left| \alpha (-\tau )\right\rangle \rightarrow
\nonumber \\
&&a_{1}a_{2}\left| \alpha (2\tau )\right\rangle
+(a_{1}b_{2}+b_{1}a_{2})\left| \alpha (0)\right\rangle +b_{1}b_{2}\left|
\alpha (-2\tau )\right\rangle \rightarrow  \nonumber \\
&&a_{1}a_{2}a_{3}\left| \alpha (3\tau )\right\rangle
+(a_{1}a_{2}b_{3}+a_{1}b_{2}a_{3}+b_{1}a_{2}a_{3})\left| \alpha (\tau
)\right\rangle +  \nonumber \\
&&(a_{1}b_{2}b_{3}+b_{1}a_{2}b_{3}+b_{1}b_{2}a_{3})\left| \alpha (-\tau
)\right\rangle +b_{1}b_{2}b_{3}\left| \alpha (-3\tau )\right\rangle .
\label{ev2}
\end{eqnarray}
In order to adjust the coefficients in (\ref{ev2}) to reproduce the phases
of $\alpha _{k}$ in Eq. (\ref{psin}), we need first to attribute values to
the duration of each evolution pulse and to the Lamb-Dicke parameter, $\tau =%
\frac{\pi }{(M+1)\bar{ \Omega}},\qquad \eta =\frac{1}{\sqrt{2(M+1)}}. $

For determining the time intervals $t_{l}^{\prime }$ of each rotating pulse
we set each coefficient in (\ref{ev2}) equal to $1$, so we need to solve a
system of algebraic equations. Calling $z_{l}=b_{l}/a_{l}$, the system of
equations is
\[
z_{1}z_{2}z_{3} =1\, ,\quad z_{1}+z_{2}+z_{3} = 1 \, , \quad
z_{1}z_{2}+z_{2}z_{3}+z_{1}z_{3} =1\,,
\]
whose solution are the roots of the polynomial equation $z^{3}-z^{2}+z-1=0$.
In general, after $M$ successful cycles the solution of the equation
\begin{equation}
\sum_{k=0}^{M}(-1)^{M-k}z^{k}=0,
\end{equation}
are the roots $z_{l}=e^{i\pi +2\pi il/(M+1)},l=1,2,...,M$. But since $%
z_{l}=e^{2i\Lambda t_{l}^{\prime }}$, so the $l-$th pulse time interval must
be $t_{l}^{\prime }=\left( \frac{l}{M+1}+\frac{1}{2}\right) \frac{\pi }{%
\Lambda } $ . The probability to produce such a state is
\begin{eqnarray}
P_{\uparrow }\left( M\tau +\sum_{l=1}^{N}t_{l}^{\prime }\right) &=&\frac{1}{
2^{2M}}\left\{ M+1+2\sum_{k=1}^{M}k\exp \left[ -2r^{2}\sin ^{2}\left( \pi
k/\left( M+1\right) \right) \right. \right.  \nonumber \\
&\times& \left. \left. \cos \left[ r^{2}\sin \left( 2\pi k/\left( M+1\right)
\right) \right] \right] \right\} ,  \label{pb2}
\end{eqnarray}
and the total time of pulses is $T=M\tau +\sum_{l=1}^{M}\left( \frac{l}{M+1}+%
\frac{1}{2}\right) \frac{\pi }{ \Lambda }=M\left( \frac{2\pi }{ \Omega }+%
\frac{\pi }{\Lambda }\right)$. One verifies that for a large $M$, $%
P_{\uparrow }\left( M\tau +\sum_{l=1}^{M}t_{l}^{\prime }\right) $ is very
small and $T$ is very large. An example: for engineering $M+1=16$ superposed
states one needs $15$ cycles, with \mbox{$ P_{\uparrow
}(t_{1}+...+t_{15})\approx 1/2^{15}$}; the total duration of the pulses is $%
T=15\left( 2\pi / \Omega +\pi /\Lambda \right)$, exceeding the total time of
the previous scheme by $15\pi /\Lambda $ which is large because experimental
conditions imply $\Lambda \ll \Omega $ since $\eta _{r}\lesssim \eta $. Thus
it is highly unlikely to generate a Fock state for $N>4$ since the
probability is quite small, whereas for a small $M$ we remind that the
condition $\left( er^{2}/\left( M+1\right) \right) ^{M+1}\gg 1$ must be
fulfilled. %

\subsection{Third scheme - $N=2^{M}$ odd circular states}

%
To construct the superposition of $2^{M}$ coherent states of odd parity (\ref
{psin2}) we have to combine the two previous schemes, since now we need one
rotating pulse, one evolution pulse and one measurement per cycle. A pair of
lasers in-phase is used to produce a rotating pulse, such that $%
H_{r}=-\Lambda \sigma _{x}$.

The evolving and rotating pulses have duration $t_{k}=\pi /\left( 2^{k}\bar{
\Omega}\right) $ \ and $t_{l}^{\prime }=\pi /\left( 2^{l}\Lambda \right) $,
respectively, and $\eta _{r}^{2}\Lambda \ll \eta ^{2} \Omega $. After $M$
successful cycles the ion state becomes
\begin{equation}
\left| \tilde{\Psi}^{\prime }\left(
\sum_{k=1}^{M}t_{k}+\sum_{l=1}^{M}t_{l}^{\prime }\right) \right\rangle =%
\frac{{\cal N}_{M}}{2^{M}}\left| \tilde{\Psi}_{2^{M}}\right\rangle \otimes
\left| \uparrow \right\rangle ,
\end{equation}
with probability
\begin{eqnarray}
P_{\uparrow }\left( \sum_{k=1}^{M}t_{k}+\sum_{l=1}^{M}t_{l}^{\prime }\right)
&=& \frac{1}{2^{M}}\left\{ 1+2^{-(M-1)}\sum_{k=1}^{2^{M}-1}k\exp \left[
-2r^{2}\sin ^{2}\left( \frac{ \pi k}{2^{M}}\right) \right] \right.  \nonumber
\\
&\times & \left. \cos \left[ \frac{2\pi k}{2^{M}}+r^{2}\sin \left( \frac{%
2\pi k}{2^{M}}\right) \right] \right\} .
\end{eqnarray}
and the total duration of the pulses is $T=\left( 2^M -1\right) \left( \frac{%
2 \pi}{ \Omega}+\frac{\pi } {2^M \Lambda} \right)$. We remind that $\left(
r^{2}e/2^{M+2}\right) ^{2^{M}}\ll 1$ is the necessary condition for
generating approximately a Fock state $\left| 2^{M}-1\right\rangle $ from
odd parity circular state ($2^{M}$ coherent states).

An example: for $M=4$ the probability is maximum, $P_{\uparrow }\left(
\sum_{k=1}^{4}t_{k}+\sum_{l=1}^{4}t_{l}^{\prime }\right) \approx 0.1 $, for $%
r=4$, thus generating approximately the Fock number state $\left|
15\right\rangle $. However, the total time involved in the engineering of
the state $\left| \tilde{\Psi}^{\prime }\left(
\sum_{k=1}^{4}t_{k}+\sum_{l=1}^{4}t_{l}^{\prime }\right) \right\rangle
\approx |15 \rangle $ is $T=15\left( 2\pi / \Omega +\pi /16\Lambda \right) $%
, which is larger than in the first scheme by $15\pi /16\Lambda $, showing
that it takes more pulses time to generate an odd Fock number state than an
even one. %

\section{Conclusions}

%
We have analyzed three schemes to engineer particular circular states for
the vibrational mode of a trapped ion. We have considered the ion
interacting with two laser beams in a stimulated Raman configuration and
have selected the effective laser frequency in resonance with the ion
transition electronic frequency, which resulted into an effective
interaction of type-Kerr in the ion-laser system. After preparing the system
in a particular initial state $\left( \left| \alpha _{0}\uparrow
\right\rangle \right) $ we have considered $M$ operations where each
successful cycle results in a new (extended) circular state for the
vibrational motion. We have calculated the total time of pulses necessary
for each kind of circular state and the probability to produce it. We have
shown the possibility to construct Fock states and two-Fock states
superpositions out from the circular states and verified that it takes much
more pulses time to generate an odd circular state $|2^M -1\rangle$ than the
even $|2^M\rangle$.

The measurement process occurs when the laser is coupled to the $|d\rangle
\leftrightarrow |\downarrow \rangle$ transition, where the width of the $%
|d\rangle$ level is $\Gamma$ and $\Gamma/2\pi \approx 20$ MHz ($2\pi/\Gamma
\approx 0.05 \mu $ s), the duration of the pulse $T^{\prime}$ must be
sufficiently long to allow the emission of at least one photon with high
probability. The non-observation of a fluorescence photon during time $%
T^{\prime}$ is a measurement, meaning that the internal electronic state is
in level $|\uparrow \rangle $, thus collapsing the state, and so the process
of the other lasers pulses can proceed. According to \cite{dav96} $%
T^{\prime}\ll 2 \mu $s, assuming a $T^{\prime}\approx 0.2 \mu$s, the time $%
NT^{\prime}$ , where $N$ is the number of cycles of pulses, is to be added
to the pulses total time for producing successfully a Fock number state. At
much it will represent about or less than $10\%$ of the total time. %
\acknowledgments{SSM thanks the agency CNPq (Bras\'{i}lia, Brazil) for
partial financial support}
%


\begin{references}
\bibitem{yur86}  B. Yurke and D. Stoler, Phys. Rev. Lett. {\bf 57}, 13
(1986).

\bibitem{dav92}  M. Brune, S. Haroche, J. M. Raimond, L. Davidovich and N.
Zagury, Phys. Rev. A {\bf 45}, 5193 (1992).

\bibitem{vas93}  K. Vogel, V. M. Akulin and W. P. Schleich, Phys. Rev. Lett.
{\bf 71}, 1816 (1993).

\bibitem{jan95}  J. Janszky, P. Domokos, S. Szab\'{o} and P. Adam, Phys.
Rev. A {\bf 51}, 4191 (1995).

\bibitem{jan96}  S. Szab\'o, P. Adam, J. Jansky and P. Domokos, Phys. Rev. A
{\bf 53}, 2698 (1996).

\bibitem{zg97}  S. B. Zheng and G. C. Guo, Quant. Semiclass. Opt. {\bf 9},
L45 (1997).

\bibitem{gerry97}  C. C. Gerry, Phys. Rev. A {\bf 55}, 2478 (1997).

\bibitem{laweber96}  C. K. Law and J. H. Eberly, Phys. Rev. Lett. {\bf 76},
1055 (1996).

\bibitem{ragi00}  R. Ragi, B. Baseia and S. S. Mizrahi, {\em J. Opt. B} {\bf %
2}, 306 (2000).

\bibitem{meek96}  D. M. Meekhof, C. Monroe, B.E. King, W. M. Itano and D. J.
Wineland, Phys. Rev. Lett. {\bf 76}, 1796 (1996).

\bibitem{resf}  F. Diedrich, J. C. Berquist, W. M. Itano and D. J. Wineland,
Phys. Rev. Lett. {\bf 62}, 403 (1989); \newline
W. Paul, Rev. Mod. Phys. {\bf 62}, 531 (1990); \newline
M. G. Raizen, J. M. Gilligan, J. C. Berquist, W. M. Itano and D. J.
Wineland, Phys. Rev. A {\bf 45}, 6493 (1992); \newline
C. A. Blockley and D. F. Walls Phys. Rev. A {\bf 47}, 2115 (1993); \newline
C. Monroe, D. M. Meekhof, B. E. King, S. R. Jefferts, W. M. Itano and D. J.
Wineland, Phys. Rev. Lett. {\bf 75}, 4011 (1995);\newline
D.J. Wineland, C. Monroe, W. M. Itano, D. Leibfried, B. E. King and D. M.
Meekhof, J. Res. NIST {\bf 103}, 259 (1998)

\bibitem{roos99}  Ch. Roos, Th. Zeiger, H. Rohdle, H. C. N\"{a}gerl, J.
Eschnner, D. Libfried, F. Schmidt-Kaler and R. Blatt, Phys. Rev. Lett. {\bf %
83} 4713, (1999).

\bibitem{mv196}  R. L. de Matos Filho and W. Vogel, Phys. Rev. Lett. {\bf 76}
608, (1996); {\bf 76}, 4520 (1996).

\bibitem{dav96}  L. Davidovich, M. Orsag and N. Zagury, Phys. Rev. A {\bf 54}%
, 5118 (1996).

\bibitem{kneerlaw98}  B. Kneer and C. K. Law, Phys. Rev. A {\bf 57}, 2096
(1998).

\bibitem{cir93}  J. I. Cirac, R. Blatt, A. S. Parkins and P. Zoller, Phys.
Rev. Lett. {\bf 70} 762 1993;

\bibitem{drobny98}  J. Steinbach, J. Twanley and P. L. Knight, Phys. Rev. A
{\bf 56}, 4815 (1997);\newline
G. Drobn\'{y}, B. Hladk\'{y} and V. Bu\v {z}ek, Phys. Rev. A {\bf 58}, 2481
(1998).

\bibitem{bas99}  J. M. C. Malbouisson, B. Baseia {\em Journ. Mod. Opt.} {\bf %
46}, 2015 (1999).

\bibitem{zg98}  S. B. Zheng, G. C. Guo, Eur. Phys. J. D {\bf 1}, 105 (1998).

\bibitem{wagner00}  W.D. Jos\'e and S. S. Mizrahi, J. Opt. B {\bf 2}, 306
(2000).

\bibitem{moya99}  H. Moya-Cessa, S. Wallentowitz and W. Vogel, Phys. Rev. A
{\bf 59}, 2920 (1999).

\bibitem{jan93}  J. Janszky, P. Domokos and P. Adam, Phys. Rev. A {\bf 48},
2213 (1993); \newline
P. Domokos, J. Janszky, P. Adam and T. Larsen, Quantum Opt. {\bf 6}, 187
(1994).

\bibitem{gag95}  M. J. Gagen, Phys. Rev. A {\bf 51}, 2715 (1995).

\bibitem{mv98}  R. L. de Matos Filho and W. Vogel, Phys. Rev. A {\bf 58}
R1661, (1998).
\end{references}
\end{document}